\documentclass[12pt]{iopart}

\usepackage{iopams}  
\usepackage{graphicx}

\newcommand{\bra}[1]{\ensuremath{\langle#1|}}

\newcommand{\ket}[1]{\ensuremath{|#1\rangle}}

\begin{document}

\title[Generation of entangled photon pairs in the bad cavity limit]{Generation of entangled photon pairs in optical cavity-QED: Operating in the bad cavity limit}

\author{R. Garc\'ia-Maraver, K. Eckert, R. Corbal\'an and J. Mompart}

\address{Departament de F\'{i}sica, Universitat Aut\`{o}noma de Barcelona, E-08193 Bellaterra, Spain}
\ead{jordi.mompart@uab.cat}
\begin{abstract}
We propose an optical cavity-QED scheme for the deterministic generation of polarization entangled photon pairs  that operates with high fidelity even in the bad cavity limit. The scheme is based on the interaction of an excited four-level atom with two empty optical cavity modes via an adiabatic passage process. Monte-Carlo wave function simulations are used to evaluate the fidelity of the cavity-QED source and its entanglement capability in the presence of decoherence. In the bad cavity limit, fidelities close to one are predicted for state-of-the-art experimental parameter values.
\end{abstract}


\pacs{42.50.Pq, 42.50.-p, 42.50.Dv}

\vspace{2pc}
\noindent{\it Keywords}: Cavity-Quantum Electrodynamics, 
Polarization-Entangled Photon Source, 
Stimulated Raman Adiabatic Passage

\submitto{\JPB}

\maketitle

\section{Introduction}
Entanglement is a quantum correlation that appears in composite systems and
constitutes one of the main resources in quantum information \cite{QI}. In optics, 
parametric down conversion (PDC) in a non-linear
crystal \cite{QCE1,QCE2,QCE3} is the standard technique to generate entangled photon pairs. However, the statistics of the photon number and time distributions follows,
essentially, a Poissonian law that severely restricts the range of practical applications of entangled photon sources based on PDC, e.g., for some quantum cryptography protocols \cite{Ek91}. 

In this paper, we propose a cavity Quantum Electrodynamics \cite{QED1,QED2,QED3,QED4,QED5,QED6,QED7} (cavity-QED) implementation to deterministically generate polarization-entangled photons pairs that presents three important features from a practical point of view: (i) it operates with high fidelity even in the bad cavity limit; (ii) it is very robust under fluctuations of the system parameters since it is based on adiabatically following an energy eigenstate; and (iii) the initial field state of the system is the simplest in optical cavity-QED, namely the vacuum state for all cavity modes.

Recently there have been several cavity-QED proposals for generating entangled photon pairs coupling a single atom to two e.m. modes of a single optical cavity or even by means of two optical cavities \cite{twomodescQED,twocavitiescQED1,twocavitiescQED2,twocavitiescQED3}. Common to all these cavity-QED implementations is the requirement to operate in the strong coupling regime while the work that we present here is, to our knowledge, the first cavity-QED proposal for generating entangled photon pairs that operates even in the bad cavity limit, i.e., when the lifetime of the photon in the cavity, given by the transmission rate of the cavity mirrors, is much smaller that the typical time that the vaccum modes of the cavity need to produce a single quantum Rabi oscillation. Note also that a single-atom source in a cavity-QED setup has been considered for engineering entangled many-photon pulses consisting of a sequence of non-overlapping one-photon wave packets \cite{Cirac1}. Notably, this sequential generation has been completely characterized showing that the attainable states correspond to the hierarchy of the so-called matrix-product states \cite{Cirac2}. 

The cavity-QED proposal that we discuss here to deterministically generated polarization entangled photon pairs is based on adiabatically following an energy eigenstate of the complete system atom-cavity modes that initially corresponds to an excited atom plus to empty cavity modes and, eventually, to the atom in its internal ground state and the two-photons entangled in their polarization degree of freedom. In Section 2, we introduce the physical scheme under investigation and derive the corresponding Hamiltonian. The coherent dynamics yielding the entangling protocol is discussed in Section 3. In section 4, we investigate the role of decoherence, i.e., spontaneous atomic decay and photon detection, by means of the Monte Carlo Wave Function approach. Section 5 is devoted to the characterization of the cavity-QED source entanglement capability while some practical considerations are briefly discussed in Section 6. Finally, in Section 7 we summarize the main results of the paper.

\section{Physical framework}
The system under investigation, sketched in Fig.~1(a), is composed of two longitudinal cavity modes $\omega_1$ and $\omega_2$, presenting polarization degeneracy, and two atomic transitions $F=0\leftrightarrow F'=1$ and $F'=1'\leftrightarrow F''=0''$ in a ladder configuration. As usual, $F$ represents total angular momentum and $m_{F}$ its projection along the quantization axis. In what follows we will use the notation $\ket{F_{m_F}}$ for the atomic state. The Hamiltonian of the system is composed of the free Hamiltonian of the atom and the e.m.~cavity modes $H_{\rm{atom}}$ and $H_{\rm{cav}}$, respectively, and the interaction Hamiltonian in the rotating wave approximation $H_{\rm{I}}$ ($\hbar = 1$), i.e.,:
\begin{eqnarray}
H_{T}=H_{\rm{atom}}+H_{\rm{cav}}+H_{\rm{I}}\hspace{2.2cm}\\
H_{\rm{atom}}= \omega_{2-}\left|1'_{-1}\right\rangle \left\langle 1'_{-1}\right|
+\omega_{2+}\left|1'_{1}\right\rangle \left\langle 1'_{1}\right|
+\left(\omega_{1+} + \omega_{2-}\right)\left|0_{0}\right\rangle \left\langle 0_{0}\right| \\
H_{\rm{cav}}=\sum_{i=1,2} \sum_{\alpha=+,-} \omega_{i\alpha}\left(a^{\dag}_{i\alpha} a_{i\alpha}  \right)\hspace{2.1cm}\\
\label{Hamiltonian}H_{\rm{I}}=\sum_{i=1,2} \sum_{\alpha=+,-} g_{i\alpha}(t)
\left({ a^{\dag}_{i\alpha} S_{i\alpha} + a_{i\alpha} S^{\dag}_{i\alpha}} \right),\hspace{1.2cm}
\end{eqnarray}
where $i=1,2$ denotes the two longitudinal modes and $\alpha=\pm$ referes to the two circular orthogonal polarizations. The energies of the atomic states are given as a function of the electric dipole transition frequencies $\omega_{i\pm}$ $(i=1,2)$, with state $\ket{0''_{0}}$ being the zero of energies. $g_{i\alpha}(t)$ is the corresponding time-dependent vacuum Rabi frequency for each polarization mode. $a^{\dag}_{i\pm}$ ($a_{i\pm}$) is the photon creation (annihilation) operator for each mode, and $S_{1+} = \ket{1'_{-1}}\bra{0_0}$, $S_{1-} = \ket{1'_{1}}\bra{0_0}$, $S_{2+} = \ket{0''_0}\bra{1'_1}$, $S_{2-} = \ket{0''_0}\bra{1'_{-1}}$ are atomic lowering operators. Detunings are defined as $\Delta_{1\pm}=\omega_{1}-\omega_{1\pm}$ and $\Delta_{2\pm}=\omega_{2}-\omega_{2\pm}$.
\begin{figure*}[t]
	\centering
		\includegraphics[width=0.29\textwidth]{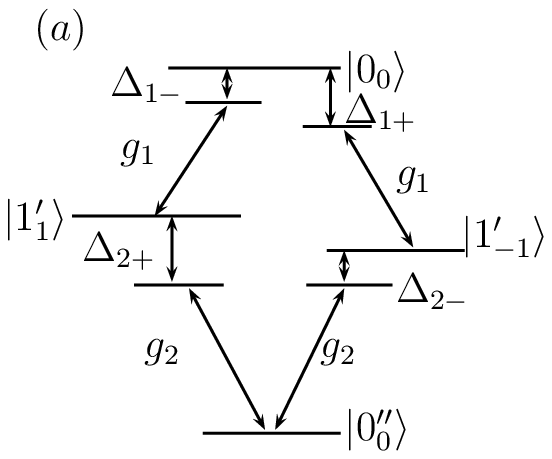}
		\includegraphics[width=0.69\textwidth]{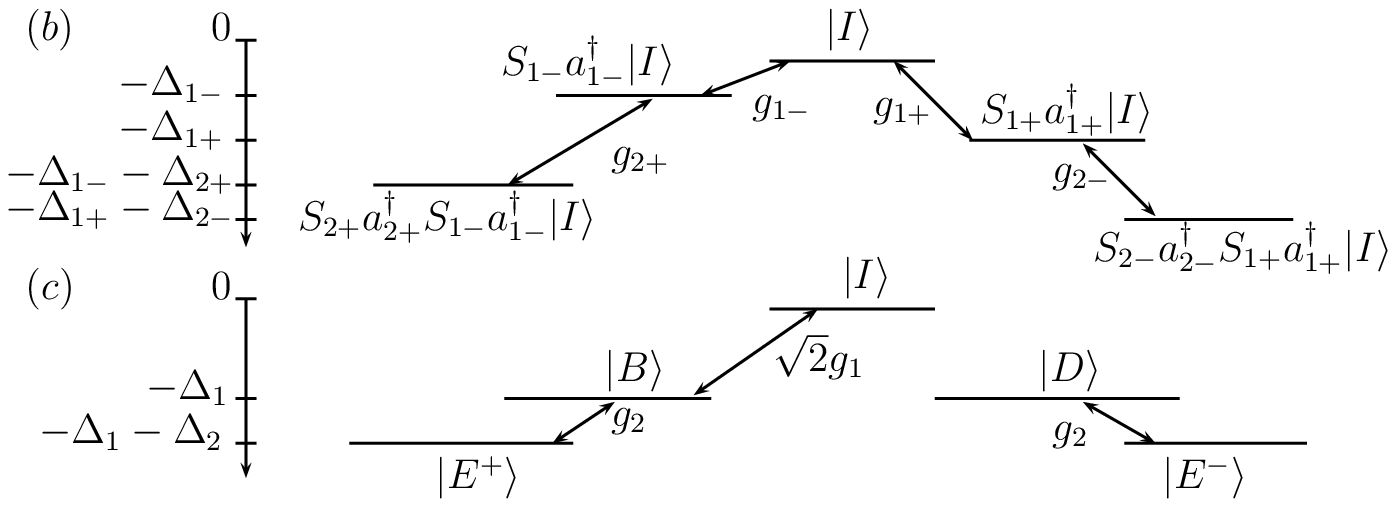}
	\label{fig:nueva}
		\caption{(a)	Four-level atomic system:  $\ket{F_{m_F}}$ denotes the atomic levels, $g_{i\alpha}$ $(i=1,2)$ $(\alpha=\pm)$ account for the vacuum Rabi frequencies of each polarization mode, and $\Delta_{i\alpha}$ $(\alpha =\pm)$ are the cavity detunings from the corresponding atomic transitions. (b) Manifold of atom-field states involving the initial state   $\ket{I}\equiv\ket{0_{0}}\otimes \ket{\Omega}$. (c) The same manifold as in (b) in the interaction picture and under the Raman resonance condition $\Delta_{1+}=\Delta_{1-}$ and $\Delta_{2+}=\Delta_{2-}$, and $g_{i+}=g_{i-}(=g_{i})$. For the definition of $\ket{B}$, $\ket{D}$, and $\ket{E^{\pm}}$ see Eqs. (\ref{rty})-(\ref{er}).}
\end{figure*}
For the sake of simplicity, we will assume equal vacuum Rabi frequencies for the two polarization states of each longitudinal cavity mode , i.e., $g_{i+}=g_{i-}=g_{i}$.

\section{Entangling mechanism}
We take the initial state of the system to be $\ket{\psi(t=0)}=\ket{0_{0}} \otimes \ket{\Omega}$ with $\ket{\Omega} \equiv  \ket{\Omega_1} \otimes \ket{\Omega_2}$, being $\ket{\Omega}_i$ the vacuum state of mode $i=1,2$.
 The coherent evolution of the system will remain in the space spanned by the five states of the manifold shown in Fig.~1(b). This evolution is, in general, very involved except for the Raman resonant case, $\Delta_{1+}=\Delta_{1-}(\equiv\Delta_{1})$, and $\Delta_{2+}=\Delta_{2-}(\equiv\Delta_{2})$ where the interactions can be reduced to that of a three-level system. In order to show this, we will consider the following basis:
\begin{eqnarray}
\label{ty}\ket{I}&\equiv&\ket{0_{0}} \otimes \ket{\Omega}, \\
\label{rty}\sqrt{2} \ket{B}&\equiv&\left( S_{1+} a^{\dag}_{1+} + S_{1-} a^{\dag}_{1-} \right) \ket{I}, \\
\sqrt{2} \ket{D}& \equiv & \left( S_{1+} a^{\dag}_{1+} - S_{1-} a^{\dag}_{1-} \right) \ket{I}, \\
\label{er}\sqrt{2} \ket{E^\pm}& \equiv & \left( S_{2-}a^{\dag}_{2-}S_{1+}a^{\dag}_{1+}  \pm  S_{2+}a^{\dag}_{2+}S_{1-}a^{\dag}_{1-} \right) \ket{I}.
\end{eqnarray}
$\ket{B}$ and $\ket{D}$ are the so-called bright and dark state \cite{Ari} combinations of the atomic states $\ket{1_{1}}$ and $\ket{1_{-1}}$ and the two circularly polarized states of mode $\omega_1$. $\ket{E^{\pm}}$ correspond to the Bell states of two orthogonally polarized photons, one in each longitudinal mode, with the atomic state $\ket{0''_{0}}$ factorising. Under the Raman resonance condition mentioned above, it is straightforward to check that the interaction picture Hamiltonian of the system, $H'_{\rm{I}}$, has the following off-diagonal matrix elements:
\begin{eqnarray}
\bra{D}H'_{\rm{I}}\ket{I}&=&\bra{D}H'_{\rm{I}}\ket{E^+}=\bra{B}H'_{\rm{I}}\ket{E^-}=0,   \\
\bra{B}H'_{\rm{I}}\ket{I}&=&\sqrt{2}g_1 e^{-i\Delta_{1} t},  \\
\bra{B}H'_{\rm{I}}\ket{E^+}&=&\bra{D}H'_{\rm{I}}\ket{E^-}=g_2 e^{i\Delta_{2} t}. 
\end{eqnarray}
The coupling chain thus reduces to that of the three-level system shown in Fig. 1(c): $\ket{I}\leftrightarrow\ket{B}\leftrightarrow\ket{E^{+}}$. 

Under the two-photon resonance condition $\Delta_{1}=-\Delta_{2}$, one of the energy eigenstates of this three-level system is 
\begin{equation}
\ket{\Lambda (\theta)}=\cos{\theta}\ket{I} - \sin{\theta} \ket{E^+},\label{D}
\end{equation}
with $\tan \theta (t) \equiv \sqrt{2} g_1 (t) /g_2 (t)$.
For $\theta=0$, Eq. (\ref{D}) corresponds to the initial state of the system: $\ket{\Lambda (0)}=\ket{I}$. 
Following the STIRAP technique \cite{STIRAP,STIRAP2,STIRAP3}, the adiabatic change of $\theta$ from $0$ $(g_1=0, g_2 \neq 0)$ to $\pi/2$ $(g_1 \neq 0,g_2=0)$ transfers the population from $\ket{I}$ to the entangled state $\ket{E^+}$ without populating the intermediate state $\ket{B}$. Since $\ket{I}$ contains no cavity photons, the leakage of two photons through the cavity mirrors indicates the success in generating state $\ket{E^+}$ which, in turns, shows the possibility of operating with high fidelity even in the bad cavity limit, i.e., for $\kappa\geq g$. The STIRAP technique requires the counterintuitive interaction of the four level atom with the cavity mode $\omega_2$ and later on, and with an appropriate temporal overlapping, with mode $\omega_1$. Notably, this scheme is very robust under fluctuations of the experimental parameters provided the adiabaticity condition is satisfied during all the interaction process. 

\section{Monte Carlo Wave Function simulations}

To numerically investigate the previous proposal in the presence of dissipation, we will use next the Monte Carlo Wave Function (MCWF) formalism \cite{DCM92}. In this approach, the time evolution of the wave fucntion of a single quantum system, a so-called quantum trajectory, is calculated by integrated the time-dependent Schr\"'odinger equation using an effective non-Hermitian Hamiltonian. Incoherent process, such as spontaneous emission or photon detection, are incorporated as quantum-jumps ocurring at random times. Thus, a quantum trajectory consists of a series of coherent evolution periods separated by quantum-jumps. Notably, the MCWF formalism is equivalent to the density-matrix formalism but provides better insights into the underlying physical mechanisms \cite{QJ1,QJ2,QJ3} and accordingly we will use it here. For the system under investigation, the non-hermitian Hamiltonian is
\begin{equation}
H_{\rm{eff}}=H^{\prime}_{\rm{I}}- \sum_{i=1,2}\sum_{\alpha=+,-} \left(i{\Gamma \over 2}S_{i\alpha }^{\dag}S_{i\alpha}+ i {\kappa \over 2} a_{i\alpha}^{\dag}a_{i\alpha}\right),
\end{equation}
where $\kappa$ accounts for the cavity-photon emission rate and the eventual photon-detection, and $\Gamma$ corresponds to the spontaneous atomic decay rate. For simplicity, we have taken equal cavity and atomic decay rates for all cavity modes and for the four atomic transitions, respectively. As usual in the MCWF formalism, to calculate a single quantum trajectory a new pseudorandom number $\epsilon $ is used at each interval time $dt$ to decide whether a quantum jump occurs. In case it doesn't occur, the wavefunction has to be renormalized to ensure unitary evolution.

\begin{figure}[h!]
	\centering
		\includegraphics[width=0.43\textwidth]{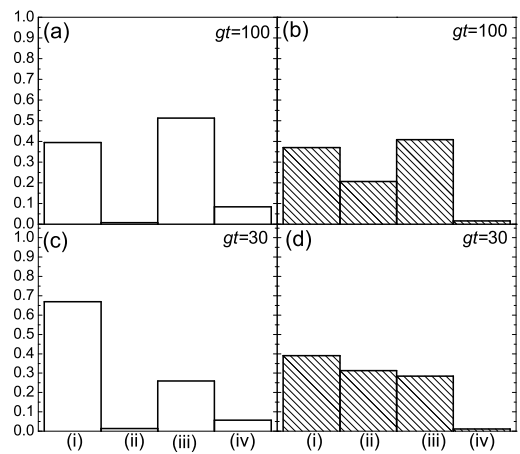}
	\caption{Probabilities for (i) generating a maximally entangled photon pair; (ii) generating two non-entangled cavity photons; (iii) emitting one cavity photon and one spontaneously emitted photon; (iv) emitting two photons by means of spontaneous emission.  Parameters are $\kappa=0.1g$ and $\Gamma=0.01g$ for (a) and (c), and $\kappa=g$ and $\Gamma=0.01g$ for (b) and (d). In all cases $\Delta_{1}=\Delta_{2}=0$. Results are obtained averaging over many Monte Carlo wave function simulations. The time duration of the entanglement procedure is $gt=100$ for (a) and (b) and $gt=30$ for (c) and (d).}
	\label{fig:barras}
\end{figure}

After averaging over many quantum trajectories, we have obtained the probabilities associated to all possible photon-emission events shown in Figs. \ref{fig:barras}(a) and \ref{fig:barras}(c) for $\kappa=0.1g$ and $\Gamma=0.01g$, and Figs. \ref{fig:barras}(b) and \ref{fig:barras}(d) for $\kappa=g$ and $\Gamma=0.01g$. In the four cases, we have considered properly overlapped temporal Gaussian profiles for the  interaction strengths.

\begin{figure}[h!]
\centering
\includegraphics[width=0.60\textwidth]{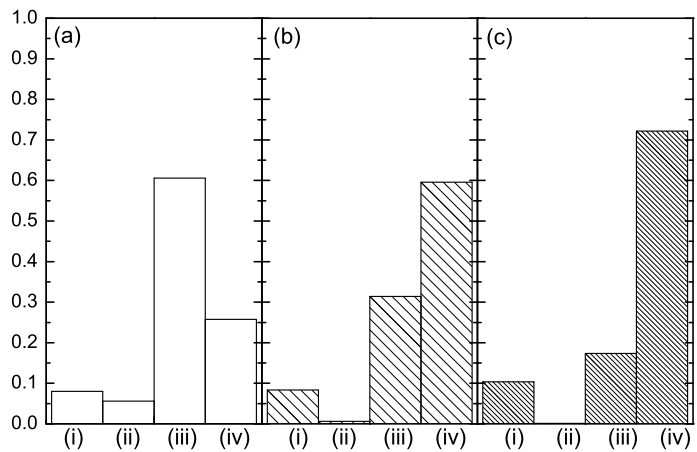}
\caption{Probabilities for events (i)-(iv). Parameters are $\Delta_{1}=-\Delta_{2}=5g$ for (a), $\Delta_{1}=-\Delta_{2}=10g$ for (b) and $\Delta_{1}=-\Delta_{2}=15g$ for (c). In all cases $\kappa=2g$ and $\Gamma=0.01g$ and the time duration of the entanglement procedure is $gt=100$. Results are obtained averaging over many Monte Carlo wave function simulations.}
	\label{fig:barrasII}
\end{figure}

\begin{figure}[h!]
\centering
\includegraphics[width=0.60\textwidth]{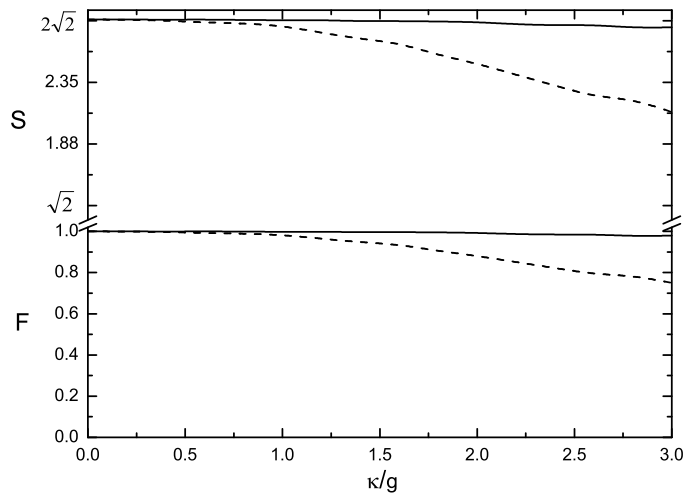}
\caption{The CHSH parameter $S$ and the Fidelity $F$ as a function of the cavity decay rate. The parameters are $\Gamma=0.01g$, $gt=100$ and $\Delta_{1}=-\Delta_{2}=7g$ for the pointed line and $\Delta_{1}=-\Delta_{2}=15g$ for the solid line.}
\label{fig:FS1}
\end{figure}

\begin{figure}[t!]
\centering
\includegraphics[width=0.40\textwidth]{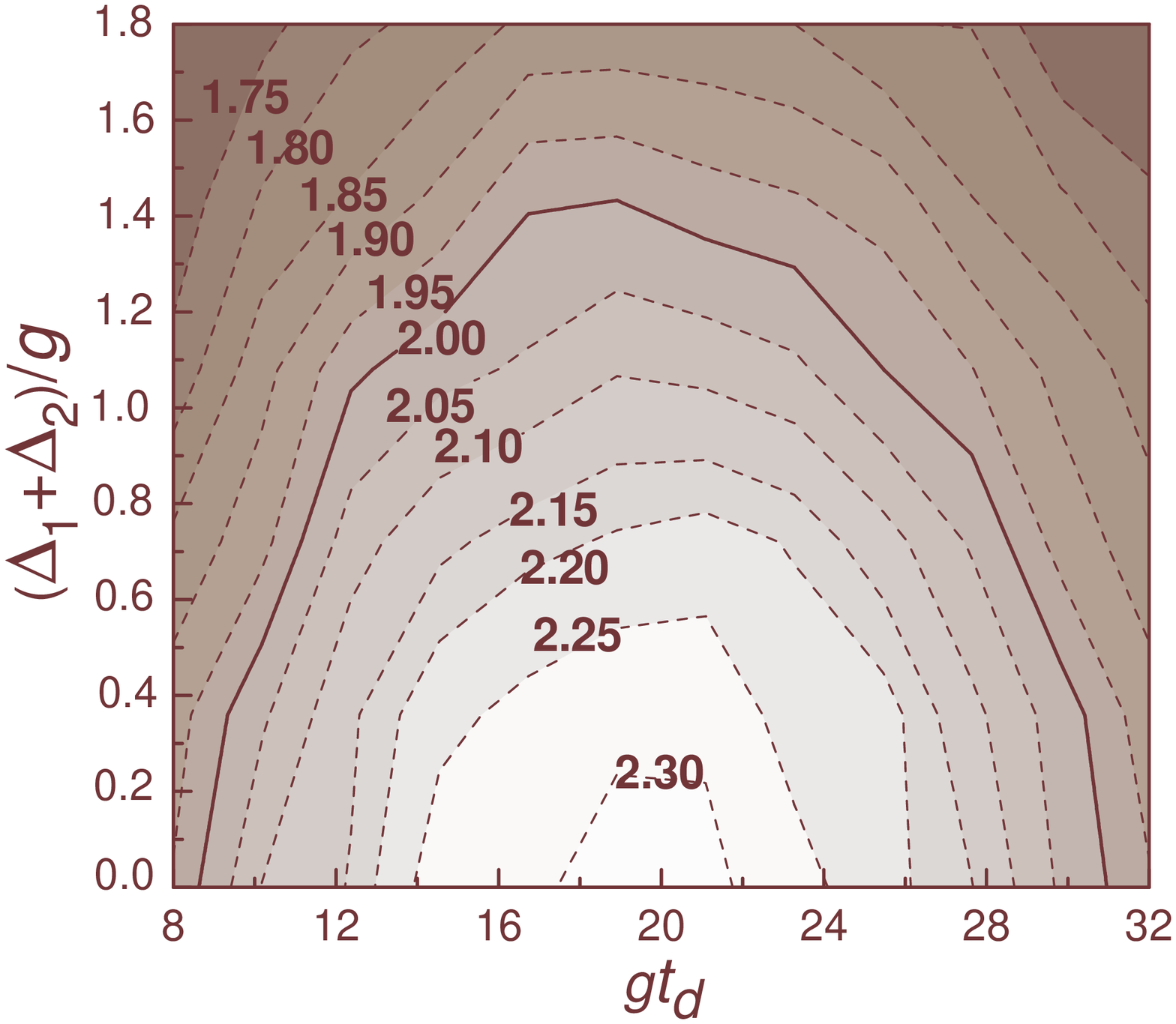}
\caption{(a) Entanglement capability $S$ as a function of the deviation from the two-photon resonace condition and the delay time $gt_{d}$ between the time-dependent coupling strengths of modes $1$ and $2$. Parameters are $\kappa=g, \Gamma=0.01g$, $\Delta_{1+}=\Delta_{1-}=\Delta_{1}$, $\Delta_{2+}=\Delta_{2-}=\Delta_{2}$ and the width (FWHM) of the coupling pulses is $gt_{w}=27$. The time duration of the entanglement procedure is $gt=100$. }
\label{fig:fig3}
\end{figure}

As it is inferred from Fig. \ref{fig:barras}(b) and \ref{fig:barras}(d), even in the bad cavity limit, the probability of generating an entangled pair of photons (corresponding to column (i)), is larger than the probability of creating a separable state of two photons emitted from the cavity (column (ii)). The later process is due to the photon-cavity decay from state $\ket{B}$, i.e., it originates from diabatic processes. By speeding up the sequence, Figs. \ref{fig:barras}(c-d), the population of $\ket{B}$ and the probability of generating two non-entangled photons will increase, while the probability of spontaneous emission processes, corresponding to columns (iii) and (iv), will reduce. In general, the time duration of the sequence will be optimized to maximize the probability given in colum (i) but also the ratio between the probabilities (i) and (ii).

The previous results can be largely improved by operating out from the single-photon resonance condition while maintining the two-photon resonance one, i.e.,  $\Delta_{1}=-\Delta_{2}\neq 0$. Thus, Fig. 3 shows the results averaging over many MCWF simulations for the following set of parameters $\kappa=2g$ and $\Gamma=0.01g$, yielding (a) $F=0.79$
for $\Delta_{1}=-\Delta_{2}=5g$, (b) $F=0.96)$ for $\Delta_{1}=-\Delta_{2}=10g$ and (c) $F=0.99$ for $\Delta_{1}=-\Delta_{2}=15g$.

\section{Characterization of the source}
By means of coincidence photodetection, one can keep only those events for which one photon is detected from each cavity mode and evaluate the fidelity $F=\bra{E^+}\rho\ket{E^+}$ where $\rho$ is the corresponding reduced density matrix. In addition, to quantify the entanglement capability of the source we have used the $S$ parameter of the CHSH inequality \cite{CHSH} ($S=\sqrt2$ for any separable state, $S=2\sqrt{2}$ for maximally entangled states, and $S\leq2$ for local hidden variable theories). Thus, for the simulations corresponding to Fig. 3 one obtains (a) $(F,S)=(0.79, 2.25)$
for $\Delta_{1}=-\Delta_{2}=5g$, (b) $(F, S)=(0.96, 2.73)$ for $\Delta_{1}=-\Delta_{2}=10g$ and (c) $(F,S)=(0.99, 2.81)$ for $\Delta_{1}=-\Delta_{2}=15g$.
For the set of parameters $\Gamma=0.01g$ and $gt=100$, Fig. 4 shows the fidelity $F$ and the entanglement capibility $S$ for $\Delta_{1}=-\Delta_{2}=7g$ (dotted line) and $\Delta_{1}=-\Delta_{2}=15g$ (solid line), as a funcion of the cavity losses $\kappa$. It is clearly seen that the cavity-QED proposal here discussed should operate with high fidelity well inside the bad cavity regime, e.g., for $\kappa \sim 3 g$, provided the the two cavity modes are far from single photon resonance but fulfill the two-photon resonance condition. 

In Fig.(\ref{fig:fig3}) the entanglement capability $S$ is shown as a function of the deviation from the two photon resonance condition and the delay time between the coupling strengths $g_{1}$ and $g_{2}$. Note that as soon as the two photon resonance condition is broken the STIRAP procedure fails and the $S$ parameter decreases. On the other hand, the $S$ parameter exhibits a robust behavior against the variation of the delay time between the two interactions as it is seen in the horizontal axis of Fig.(\ref{fig:fig3}).

\section{Some practical considerations}

A suitable atomic element with the configuration $F=0\leftrightarrow F'=1$ and $F'=1'\leftrightarrow F''=0''$ needed in our proposal, is calcium with its cascade $J=0 \rightarrow J=1 \rightarrow J=0$ transition $4p^2\,^1 S_0\rightarrow$ $4s^4 p^1 P_1 \rightarrow$ $4s^2\,^1 S_0$ at $\lambda = 551.3 \,{\rm nm}$ and $\lambda = 422.7 \, {\rm nm}$, respectively \cite{Aspect,Clauser}.
Alternative one could also use a $J=1 \rightarrow J=1 \rightarrow J=0$ configuration like the 
$7^3 S_1\rightarrow $ $6^3 P_1\rightarrow $ $6^3 S_0$ of the $^{200}$Hg \cite{Fry}. In both cases, one needs to perform first a double excitation process such that the atom enters the cavity-QED setup in its excited state. 
Note that as the initial state can decay by spontaneous emission, if one pretends to obtain a close to 100\% success probability in generating the entangled photon pair, then the spontaneous population decay rate should be much smaller than the inverse of the total time needed for the STIRAP process. In fact, our simulations (see Figs. 2 to 5) include not only cavity decay of photons through the mirrors  but also the spontaneous atomic decay from the optical transitions. For the simulations of these figures, the spontaneous atomic decay rate has been taken smaller but close to the inverse of the total time needed for the STIRAP process. However, if the spontaneous emission rate becomes on the order or much larger than the total time for the STIRAP process, then the success probability will obviously decrease. If so, one could perform  postselection of events, i.e, two photon coincidence measurements, and still achieve a fidelity close to 100\% in generating entangled photon pairs. 

Finally, note that the temporal control of the interactions, i.e., the switching on/off of the interaction of the atomic transition with the corresponding mode, can be implemented by detuning the atomic transitions from he corresponding modes via a uniform electric (or magnetic) field yielding the corresponding Stark (or Zeeman) shift, or even by the atomic light-shift induced by an external laser.

\section{Conclusion}
To sum up, we have discussed a robust and efficient cavity-QED proposal for the deterministic generation of polarization-entangled photon pairs. The complete entanglement procedure is based on a two mode STIRAP process which allows the cavity-QED source to operate with high fidelity even in the bad cavity limit. Even in this regime, the obtained fidelities yield quantum correlations between photon pairs well above from those that can be obtained from local hidden variable theories.

\ack
We acknowledge support from the Spanish Ministry of Education and Science (MEC)  and from the project "Quantum Optical Information Technologies" within the Consolider-Ingenio 2010 program under contracts FIS2005-01497 and CSD2006-00019, respectively. Support from the Catalan Government under contract SGR2005-00358 is also acknowledged. KE acknowledges support from the MEC.

\end{document}